# Femtosecond Pulse Generation via an Integrated Electro-Optic Time Lens

Mengjie Yu[1], Christian Reimer[2], David Barton[1,3], Prashanta Kharel[2], Rebecca Cheng[1], Lingyan He[2], Linbo Shao[1], Di Zhu[1], Yaowen Hu[1,4], Hannah R. Grant[5], Leif Johansson[5], Yoshitomo Okawachi[6], Alexander L. Gaeta[6,7], Mian Zhang[2], and Marko Lončar[1, *]

[1] John A. Paulson School of Engineering and Applied Sciences, Harvard University, Cambridge, MA 02138
[2] HyperLight, 501 Massachusetts Avenue, Cambridge, MA 02139
[3] Intelligence Community Postdoctoral Research Fellowship Program
[4] Department of Physics, Harvard University, Cambridge, MA 02138, USA
[5] Freedom Photonics, 41 Aero Camino, Goleta CA, USA
[6] Department of Applied Physics and Applied Mathematics, Columbia University, New York, NY 10027
[7] Department of Electrical Engineering, Columbia University, New York, NY 10027
* Corresponding author: loncar@seas.harvard.edu

**Integrated femtosecond pulse and frequency comb sources are critical components for a wide range of applications, including optical atomic clocks[1], microwave photonics[2], spectroscopy[3], optical wave synthesis[4], frequency conversion[5], communications[6], lidar[7], optical computing[8], and astronomy[9]. The leading approaches for on-chip pulse generation rely on mode locking inside microresonator with either third-order nonlinearity[10] or with semiconductor gain[11,12]. These approaches, however, are limited in noise performance, wavelength tunability and repetition rates[10,13]. Alternatively, sub-picosecond pulses can be synthesized without mode-locking, by modulating a continuous-wave (CW) single-frequency laser using a cascade of electro-optic (EO) modulators[1,14–17]. This method is particularly attractive due to its simplicity, robustness, and frequency-agility but has been realized only on a tabletop using multiple discrete EO modulators and requiring optical amplifiers (to overcome large insertion losses), microwave amplifiers, and phase shifters. Here we demonstrate a chip-scale femtosecond pulse source implemented on an integrated lithium niobate (LN) photonic platform[18], using cascaded low-loss electro-optic amplitude and phase modulators and chirped Bragg grating, forming a time-lens system[19]. The device is driven by a CW distributed feedback (DFB) chip laser and controlled by a single CW microwave source without the need for any stabilization or locking. We measure femtosecond pulse trains (520 fs duration) with a 30-GHz repetition rate, flat-top optical spectra with a 10-dB optical bandwidth of 12.6 nm, individual comb-line powers above 0.1 milliwatt, and pulse energies of 0.54 picojoule. Our results represent a tunable, robust and low-cost integrated pulsed light source with CW-to-pulse conversion efficiencies an order of magnitude higher than achieved with previous integrated sources. Our pulse generator can find applications from ultrafast optical measurement[19,20] to networks of distributed quantum computers[21,22].**

The ability to generate ultrashort, broadband and high-peak-power optical pulses on‐chip has been a long-sought-after goal. The rapid advancement of low-loss nanophotonic waveguides has reduced the pulse energy for achieving nonlinear spectral broadening across octaves of bandwidth to sub-picojoules via supercontinuum generation[23]. However, all demonstrations to date rely on a table-top pulse laser source which increases the system complexity, size, and cost, and thus hinders practical applications. In addition, optical pulses can be generated via microresonator frequency comb sources (for short, microcombs) through coupling a continuous-wave laser into a high-quality-factor microresonator[10]. However, microcombs are limited by their low efficiencies (ususally < 2%), comb-line power and high repetition rates, resulting in only tens of femtojoule pulse energies[10,24,25]. An alternative approach, based on compact and electrically pumped on-chip semiconductor modelocked lasers[11,13], still faces significant challenges including available pulse energy, pulse width, and optical and radio-frequency (RF) noise. Therefore, high-power and controllable integrated pulse sources are still missing, presenting a major road block to achieve fully integrated nonlinear photonic circuits.

A well-known and widely utilized table-top approach for synthesizing optical pulses uses non-resonant EO modulation of a CW laser light[14,26]. Directly driven by an electronic synthesizier, such pulse generators deliever flat, high power and coherent optical frequency combs with up to tens of GHz repetition rates. First proposed in the 1960's[16], EO comb generators have attracted significant interests in the recent decade due to their great flexibility in operating wavelengths and repetition rates as well as exceptional spectral flatness[2,27,28]. The pulse synthesis using cascaded amplitude and phase modulation can be described as a time lens approach[17], a temporal analogue of a conventional spatial lens (Fig. 1a). While a spatial lens system can focus a collimated laser beam to a spot with a small waist size on its focal plane, a time lens system can compress CW light to a short pulse at a proper dispersive focal length. This space-time duality originates from the mathematical analogy between the diffraction equation in the spatial domain and the dispersion equation in the temporal domain[19]. Three key components essential to building such a spatial/temporal lens system are: an aperture, a lens that induces a quadratic phase term in space/time, and a diffraction/dispersion medium representing the focal length of the imaging system. In an EO time lens, shown in Fig. 1b, an amplitude modulator (AM) carves the initial flat-top pulse representing a temporal aperture; a phase modulator (PM) induces a (quasi-) quadratic temporal phase; and a dispersive medium imparts a group delay dispersion (GDD) in the frequency domain which acts as the focal distance and compresses the periodic waveform into ultrashort pulses. Since the EO modulation is periodic in time with a microwave frequency $f_{MW}$, such a system converts a continuous-wave (CW) single-frequency light to a pulse train with a repetition rate of $f_{MW}$. In the frequency domain, the output features an optical comb spectrum with a flat-top envelope, which is a result of the time-to-frequency mapping due to the Fourier transform operation of a lens[29]. An important figure of merit of a time-lens system is the minimal achievable pulse width $\tau \sim 2\ln(2) V_\pi / (\pi^2 V_{MW} \times f_{MW})$, where $V_\pi$ is the half-wave voltage of the PM, and $V_{MW}$ is the microwave voltage applied on the PM. Therefore, for ultrashort pulse generation, EO modulators used in time lens must have a low $V_\pi$ and need to be able to handle large microwave power, in addition to a high EO bandwidth.

Due to the high $V_\pi$ of commerically available EO modulators, previously demonstrated systems required 2-4 stages of phase modulators combined with a separate intensity modulator to achieve the optical bandwidth corresponding to picosecond-pulses[26,27]. Each phase modulator is driven by an RF amplifier with several Watts of power consumption and tunable RF phase shifters to align their relative phases. Further more, laser amplifiers are needed to compensate for the insertion losses and limited optical power handling capabilities of discrete components used[1,27]. As a result, the complexity and high-cost of this approach have limited its adoption in emerging applications.

Here, we adress these limitations by fully integrating an EO comb source on an LN photonic chip. The rapid development of the thin-film lithium niobate (LN) platform has resulted in low-loss nanophotonic waveguides (3 dB/m) and state-of-the-art EO efficiency for modulation, storage, and control of optical signals[18]. Figure 1c shows the LN chip, with all three building blocks – one AM (in push-pull configuration), one PM, and one dispersive waveguide, integrated into a footprint as small as 25 × 7 mm. The optical layer is fabricated using a two-layer electron-beam lithography and etching process (see Methods) on an x-cut 600-nm LN substrate. The first layer is used to define the optical slab waveguides for the AM and PM with a top width of 1500 nm and an etch depth of 300 nm. This allows for an efficient modal overlap between the microwave field and the optical field, while maintaining a low optical propagation loss of 0.3 dB/cm. The second layer creates fully etched nanostructures that form the dispersive waveguide for pulse compression as well as spot-size converters[30] for efficient coupling to a distributed feedback (DFB) laser chip (2 dB, see Methods) and a lensed fiber (3 dB). Adiabatic tapering is used to ensure low-loss transitions of < 0.1 dB between optical waveguides on different layers. Both AM and PM utilize 2-cm-long travelling-wave coplanar waveguide (CPW) electrodes with parameters optimized for RF impedance, RF loss, and microwave-optical velocity matching (see Methods). Based on an unbalanced Mach-Zehnder interferometer in a push-pull configuration, the measured EO $S_{21}$ parameter of the AM indicates a 3-dB EO bandwidth of 45 GHz and a low DC switching voltage $V_\pi$ of 1.2 V at DC (see Methods). To reduce $V_\pi$ of the PM, we developed a recycling-PM design. Here, after passing one of CPW gaps, the optical waveguide loops back and passes through the other CPW gap. This design maximizes the interaction between optical and microwave fields and effectively reduces $V_\pi$ in half. To achieve this, low insertion loss optical waveguide crossings (loss 0.3 dB/per crossing) were developed with minimal cross-talk (see Methods). The $V_\pi$ of the recycling-PM is measured to be 2 - 2.5 V in the RF frequency range from 4 to 39 GHz. Specifically, the $V_\pi$ is measured to be 2.2 V and 2.5 V at 10.075 GHz and 30.135 GHz, respectively. In comparison, the state-of-the-art integrated PM has a $V_\pi$ of 3-4 V at 10 GHz[31], while a typical commercial, discrete component high-speed PM (iXblue photonics, MPZ-LN-40) has an EO bandwidth of 33 GHz, $V_\pi$ of 8.5 V at 30 GHz, and insertion loss of 2.5 dB. We note that in the recycling-PM the lowest $V_\pi$ is achieved when looped-back optical signal is in phase with microwave drive, which results in resonant behavior with microwave free-spectral range (FSR) of 2.8 GHz and a 3-dB power bandwidth ($\sqrt{2}V_\pi$) of 1.5 GHz, while keeping the overall EO bandwidth of 45 GHz (see Methods). The on-chip insertion loss of PM (with waveguide crossing) is 2 dB. The realization of such a looped structure uniquely benefits from the small bending radius and waveguide crossing achievable on a nanophotonic platform which can support tightly confined optical modes.

Next, we show that the time-lens chip allows for generation of pulses in the femtosecond regime, while significantly reducing the RF power consumption and complexity of the control circuitry, compared to discrete-component implementation. The experimental setup is shown in Fig. 2a (for more details, see Methods). Light from a DFB laser chip is edge-coupled into the LN chip, and polarized along the polar axis of LN to excite the transverse electric (TE) mode (Fig. 2b). The AM is biased at the quadrature point and driven with a peak-peak voltage $V_{pp}$ equal to the $V_\pi$ of the AM. The RF signal from a signal generator is split into two: one drives the AM, and the other drives the PM after being amplified and phase-aligned to the AM arm. The output of the chip is sent to an optical spectral analyzer and an intensity autocorrelator for spectral and temporal measurements, respectively (Fig. 2c). We first characterize the system performance without the on-chip dispersive waveguide, and instead use a single-mode fiber (SMF-28) for pulse compression, which has slightly anomalous group velocity dispersion (GVD < 0). Figure 2d,e shows the generated optical spectra with a 10-dB optical bandwidth of 6.0 nm and 12.6 nm for two different line spacings (RF drive frequencies) of 10.075 GHz and 30.135 GHz, respectively, obtained using the same chip. A total of 91 comb lines are generated at 10.075 GHz, corresponding to a modulation index of 11.2 $\pi$ radians at an RF power of 38 dBm. A total of 67 comb lines are achieved at 30.135GHz, corresponding to a modulation index of 7.8 $\pi$ radian at an RF power of 36 dBm. The modulation index is currently limited by the available power of our RF amplifiers instead of the damage threshold of our chip. The optical spectra have also shown an excellent spectral power flatness of total 35 and 25 comb lines within < 1dB power variation. We achieve the temporal focal "length" $f'$ of 7.1 ps$^2$ and 1.1 ps$^2$ at 10 and 30 GHz, which leads to a measured transform-limited pulse of 1.14-ps and 532-fs duration after pulse compression in an SMF-28 fiber with lengths of 360 and 59 meters, respectively, showing good agreement with simulation (Fig. 2d and Methods). The overall insertion loss of the chip is 8 dB including 3-dB on-chip insertion loss and 5-dB facet coupling losses. The insertion loss could be further reduced to a total of 1.3 dB with 0.5 dB/facet and 3 dB/m propagation loss via improved coupler[32] and fabrication[33]. In contrast, the state-of-the-art discrete component EO time lens gives 600-fs pulses at both 10 and 30 GHz repetition rate with 3-4 phase modulators and phase shifters with an unavoidably higher optical insertion loss of 12-18 dB and a total RF power consumption of 42 dBm (see Table 1 in Methods). Operation beyond 30 GHz RF frequency to further reduce the pulse duration is challenging due to the critical voltage-bandwidth trade-off of conventional LN modulators. However, our integrated time lens has not hit the EO bandwidth limit and could generate pulses as short as 200 fs by operating at 45 GHz assuming a 38-dBm RF power and are capable of coherent nonlinear broadening to an octave-spanning spectrum[1,34].

The integrated time-lens platform allows for optical frequency comb generation with microwave line spacings and offers compelling advantages in frequency agility, comb efficiency, and pulse energy. To illustrate this, we use the same LN chip to generate two femtosecond pulse trains at 30 GHz repetition rates simultaneously, at central wavelengths of 1543 and 1556 nm (Fig. 3a). Both pulses trains are compressed to ~520-fs duration, which is individually verified by autocorrelator measurements. The autocorrelation trace of the combined pulse trains reveals a small temporal delay of 14 ps between the pulse trains at the output of the fibers due to the non-zero GVD (see Methods). Furthermore, we demonstrate the generation of two EO combs spectrally separated as far as 19.5 THz by pumping the time-lens chip with two CW sources at 1506 and 1669 nm (Fig. 3a). Wavelength-multiplexed combs can be used to interrogate different spectral ranges in molecular spectroscopy[35], increase communication channels in data communications[2], and realize optical frequency division for microwave synthesis[36]. Importantly, the EO approach produces a comb power (or pulse energy) that scales linearly with the pump power, along with a flat spectrum. We achieve a total comb power of 16.25 mW for an on-chip pump power of 65 mW, corresponding to an on-chip comb conversion efficiency of 25% (Fig. 3b, 12.5% counting the facet loss). The comb lines spanning the entire 10-dB optical bandwidth (52 lines) has > 0.1 mW per line, with a total of 40 comb lines above 0.2 mW per line. We achieve an on-chip pulse energy of 0.54 pJ at a 30-GHz line spacing.

As a proof-of-principle of nonlinear frequency conversion, we pump a 1-km highly nonlinear fiber (HNLF) using time-lens pulses (Fig. 3c). The time-lens spectrum is broadened to 150 nm via self-phase modulation with 1.3 pJ pulse energy. In addition, four-wave mixing along with self-phase and cross-phase modulation is observed combining the pulse source with a CW probe laser into the HNLF (see Methods). Optical pulse source with a repetition rate at tens of GHz would directly link optical domain with microwave domain for self-referencing[1,37], directly pump a microresonator to resolve the outstanding challenge of bandwidth-line-number-product[37], and benefit a number of emerging applications[38] in astronomical spectrograph calibration, wavelength-division multiplexing, arbitrary wave generation and Raman spectroscopy.

One solution for on-chip dispersion management, required for pulse compression, is to replace long fiber with an integrated chirped Bragg grating[39]. Figure 4a shows the experimental set up, where we send the output of the cascaded EO modulators chip into the Bragg grating device via a circulator and characterize the pulse duration after reflection from the grating. The fin-shape Bragg grating, fabricated on a 600-nm thin film LN with an etch depth of 300 nm, is formed by alternating waveguide top width between 1.1 µm and 1.5 µm (Fig. 4b). The grating period is linearly chirped from 406.5 nm to 414.5 nm along the grating length $L$, in order to achieve a total group delay of $2n_g L/c$ over the reflection bandwidth where $n_g$ and $c$ are the group refractive index and light velocity, respectively. The grating has an ultra-low propagation loss of 0.033 dB/mm and a high reflectivity > 99.5% for the TE mode, both experimentally measured (see Methods). The reflection spectrum of the grating shows the 3-dB grating bandwidth of 1547 – 1576 nm, characterized using a tunable CW laser (Fig. 4c). The reflection spectrum allows us to determine the group delay through a simple spectroscopic measurement (see Methods). Briefly, we fit the frequency-dependent period of Fabry-Perot fringes that arise from the standing wave in the bus waveguide to determine the travel distance, and hence group delay, over the grating bandwidth. Fig. 4d shows the fitted group delay of 0.61, 1.10, and 1.60 ps/nm for gratings of length 1 mm, 1.75 mm, and 2.5 mm, respectively. The corresponding dispersion value is 610 ps/nm/m, a factor of $3.4 \times 10^4$ larger than the optical fiber (0.018 ps/nm/m). Such large group delay dispersion gives a grating length of 1.8 mm to achieve full pulse compression with a total on-chip insertion loss of 0.14 dB. The EO combs spectrum and the autocorrelator trace after reflection from the integrated grating device is recorded in Fig.4e & f. We achieve fully compressed pulse trains with a measured duration of 545fs. To the best of our knowledge, this is the first demonstration of an integrated LN chirped Bragg grating for on-chip dispersion management. Currently, the chirped grating device operates in a fully reflective mode and thus requires a circulator to minimize the losses. Fully integrated on chip solution, that combines time lens with the Bragg grating on the same chip would rely on a beam splitter instead of the circulator (that are still hard to implement on-chip), resulting in an extra 6-dB loss. One approach to mitigate this is to utilize recently demonstrated coupled Bragg grating structure [40].

An alternative approach for fully-integrated femto-second source is to replace Bragg grating with a dispersion-engineered waveguide for on-chip compression. The fabricated device is shown in Figure 1c, and illustrated in Figure 5a. Using this approach, we were able to integrate all components the time lens device on a single LN chip, including both modulators and 9.56 cm long dispersive compression waveguide. To achieve the required high dispersion, the optical mode is transfromed from a 600-nm tall rib waveguide to a 300 × 500 nm trapezoid waveguide using an adiabatic taper, see Fig. 5b. The transition loss is negligible (< 0.1 dB). The waveguide dimensions are chosen to achieve a dispersion value of – 2.15 ps/nm/m (GVD: +2770 ps$^2$/km) at the pump wavelength of 1557 nm, a factor of 10 and 120 higher than the absolute dispersion value of the rib waveguide and SMF-28 fibers, respectively (Fig. 5c). The high dispersion reduces the required waveguide length for pulse compression from 59 meters in fiber to 49 cm in a LN waveguide, which is within a practical range for low-loss nanophotonic fabrication and promises a significantly reduced footprint. However, the current propagation loss of the 300 nm thick section of the waveguide used for compression is relatively high, on the order of 1.5 dB/cm, which limits the length of the waveguide that could be used with reasonable losses to 9.56 cm long. As a result, the pulse is not fully compressed, resulting in 8-ps on-chip pulse duration (Fig. 5d, see Methods). We note that this is not a fundamental limitation, but rather the result of our specific fabrication process with increased scattering on the top waveguide surface (due to etching) and a less-confined optical mode. Ongoing efforts will further reduce the linear loss of the waveguide[41]. This result is a critical step for a fully integrated femtosecond pulse source enabled by an electro-optic time lens.

In conclusion, we demonstrate a flat-top and frequency-agile EO comb generator on a LN photonic chip and generation of 532-fs pulses at a 30-GHz repetition rate, directly pumped with a chip-scale DFB laser. Its realization is based on construction of an EO time-lens system, which is, for the first time, realized on a miniaturized optoelectronic circuit since its proposal in 1960s. The combination of low-loss and tightly-confined optical waveguides, high EO efficiency, and on-chip Bragg grating tremendously reduces the footprint, microwave and optical power consumption, and complexity of the entire time-lens system, making it compatible with low-cost wafer-scale production. Our EO comb generator offers an excellent solution for frequency-agile operation on chip with tunable repetition rates and wavelengths, without any limitation from cavity resonances or the gain bandwidth present in microcombs and modelocked lasers. Integration with an on-chip laser[42] could help realize a fully integrated pulse source. Reducing the microwave losses using segmented electrodes to 0.25 dB/cm/ $GHz^{1/2}$ can reduce the temporal focal length and increase the EO bandwidth for realizing 200-fs pulses[43]. One could also envisage an entire nonlinear photonic system where the EO-based pulse source serves as the pump for octave-spanning frequency combs, optical parametric oscillation and Raman-scattering microscopy. The EO comb generator itself can serve as a multi-wavelength source for microwave photonics, telecommunication, and calibration of astronomical spectrographs. Finally, the electro-optic lensing could enable the temporal and spectral shaping of single photon[21,22], which could find applications in building the future quantum network[44].

**Acknowledgements**
This work is supported by the Defense Advanced Research Projects Agency (DARPA) under Contracts No.HR0011-20-C-0137 and W911NF2010248, Draper grant N00014-18-C-1043, and AFOSR (FA9550-19-1-0376). Device fabrication was performed at the Harvard University Center for Nanoscale Systems. D.Z. acknowledges support by the Harvard Quantum Initiative post-doctoral fellowship. This research D. B. performed was supported by an appointment to the Intelligence Community Postdoctoral Research Fellowship Program at Harvard University, administered by Oak Ridge Institute for Science and Education through an interagency agreement between the U.S. Department of Energy and the Office of the Director of National Intelligence.

**Author contributions**
M. Y. and Y.O. conceived the idea. M. Y. designed the chip with the help of C. R., D. B., P. K., L. H. and M. Z. C. R. and L. He fabricated the devices. D. B. fabricated the grating device. M.Y. carried out the measurement and analyzed the data with the help of P. K., D. B., R. C., D.Z., and Y. O.. M. Y. performed numerical simulations with the help of Y. O.. H. G. and L. J. helped with the project. L. S. and Y. H. helped with the fabrication. M. Y. wrote the manuscript with contribution from all authors. M.L. supervised the project.

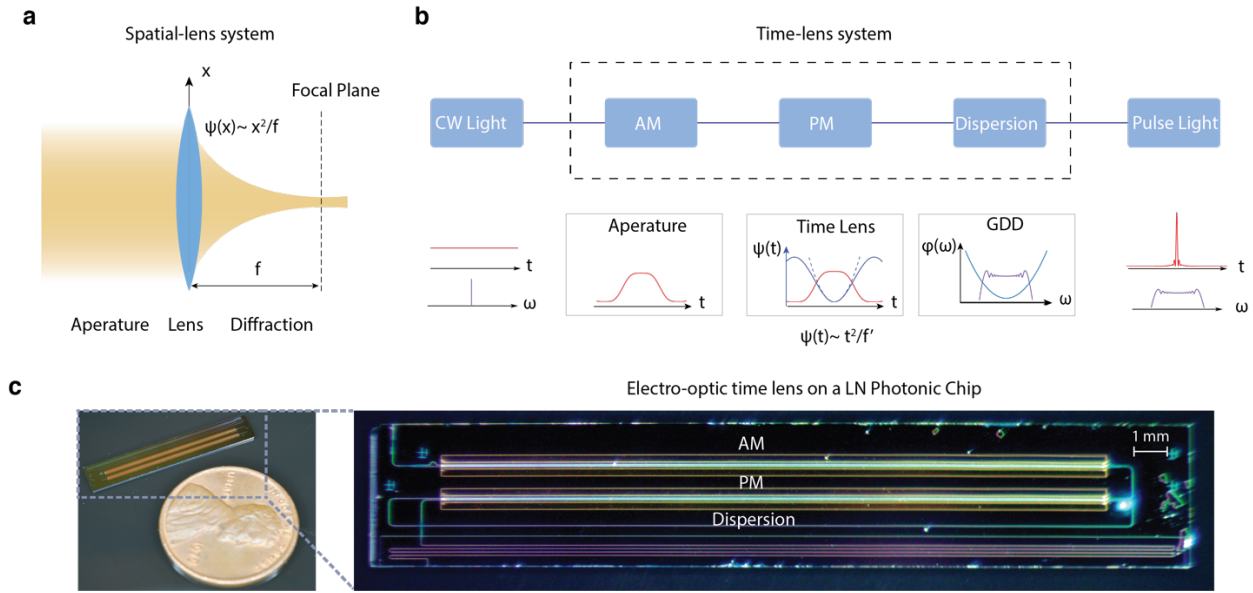

**Fig. 1 | Concept of electrical synthesis of optical pulses via a time lens system. a,** Spatial focusing via a spatial lens. A collimated beam passes a spatial lens through its aperture, undergoes diffraction in the free space, and is focused in the focal plane. A perfect lens impacts a quadratic phase profile ($\psi(x) \propto x^2/f$) in the transverse dimension, where $f$ is the focal length of the lens. **b,** Temporal focusing via an electro-optic time lens (TL). Similar to the spatial focusing in **a**, a continuous-wave (CW) light passes through a time lens system and is "focused" into a Fourier-transform-limited pulse train. The time lens system is implemented via electro-optic (EO) modulation, consisting of an amplitude modulator (AM), a phase modulator (PM) and a dispersive medium. The AM carves the CW light into the temporal aperture of the lens. The PM constructs an EO time lens to chirp the initial pulse, where an approximate quadratic phase term ($\psi(t) \sim t^2/f'$) is established at the local minima (or maxima) of a sinusoidal microwave drive, where $f'$ is the focal "length". The dispersive medium induces group delay dispersion (GDD), where different spectral components experience a time delay proportional to the derivative of the spectral phase profile $\varphi(\omega)$ with respect to frequency $\omega$. Thus, it retimes all the spectral components to form an ultrashort pulse train. This electrical synthesis of optical pulses generates a flat-top EO comb spectrum with a line spacing equal to the MW drive frequency, with its spectral bandwidth inversely proportional to the pulse duration. **c,** Optical micrograph of the integrated lithium niobate (LN) photonic chip for the time-lens implementation. Zoom-in picture (right) shows the full optoelectronic integrated circuit with a footprint of 25 × 7 mm. The photonic circuit integrates the following components on a single X-cut 600-nm LN chip: a spot-size converter, a Mach-Zehnder interferometer (MZI) -based AM, a "recycling" PM, and a highly dispersive waveguide.

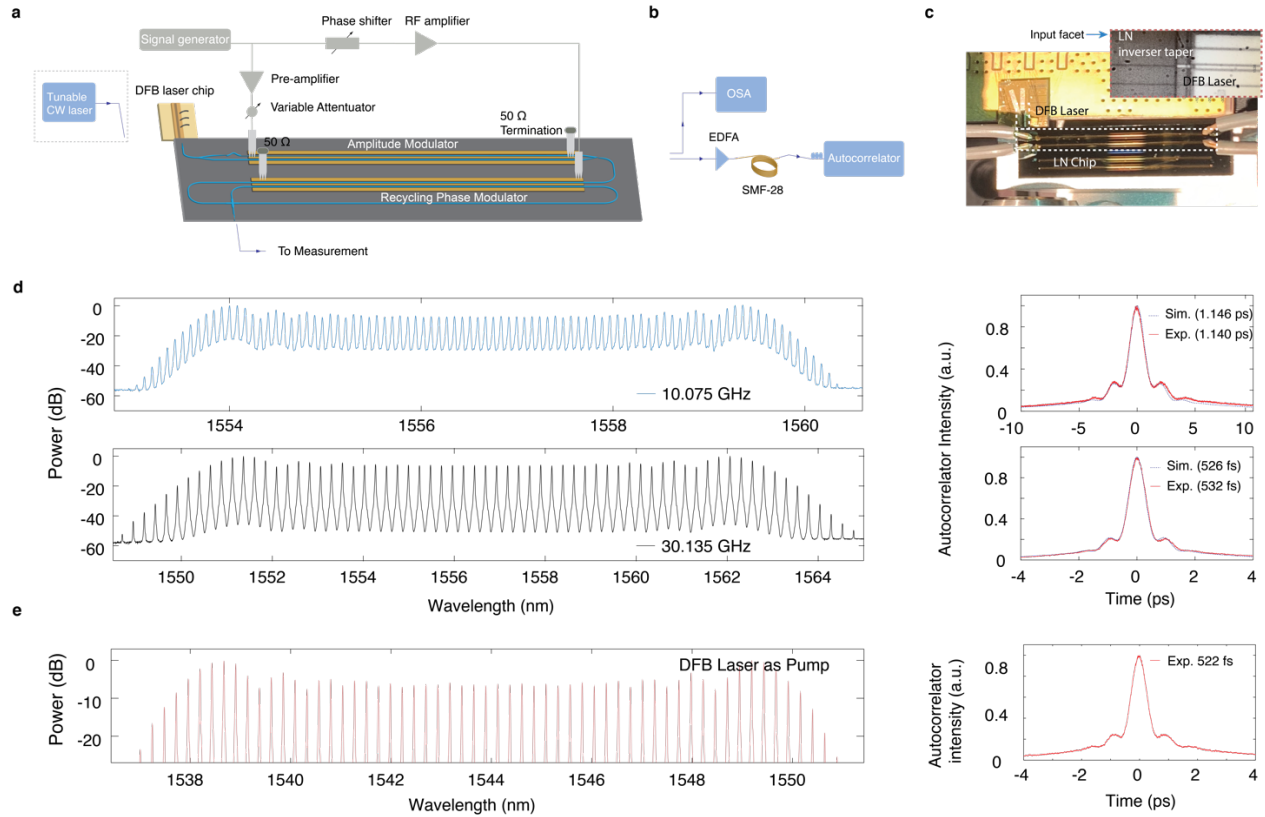

**Fig. 2 | Femtosecond pulse generator via an integrated LN-chip-based electro-optic time lens.**
**a,** Schematic of the LN chip. In this experiment, the input laser is either a tunable CW laser or a semiconductor-chip-based distributed feedback (DFB) laser. **b,** Characterization setup. The chip output is sent to an optical spectral analyzer for frequency-domain measurement or intensity autocorrelator for time-domain measurement. **c,** The optical microscope picture of the LN photonic chip with edge-coupled DFB laser. The coupling loss is 2 dB with an optimized inverse taper for edge coupling to the DFB laser output (zoom-in). **d,** The output optical spectra and temporal traces at microwave driving frequencies of 10.075 and 30.135 GHz using a tunable CW laser. Two separate EO combs are shown: At 10.075 GHz, a total of 91 comb lines are generated with a 10-dB optical bandwidth of 6.0 nm, and at 30.135 GHz, a total of 67 comb lines are generated with a 10-dB optical bandwidth of 12.6 nm. At 10.075 GHz, the TL focal "length" (referred as focal GDD) is 7.1 ps$^2$, corresponding to a PM modulation depth of 11.2$\pi$ radian. Aided with the dispersion of single-mode fiber (SMF-28) for full temporal compression, we measured the final pulse duration [full-width-half-maximum (FWHM)] to be 1.140 ps. At 30.135 GHz, the TL focal GDD is 1.1 ps$^2$, corresponding to a modulation depth of 7.8$\pi$ radian. The final pulses have a measured FWHM of 532 fs. The autocorrelator traces are in good agreement with the simulation (dotted curves). **e,** The optical spectrum with a 30-GHz line spacing and autocorrelator time trace obtained using a DFB laser pump at 1544 nm. The laser wavelength is tuned via changing the injected current, and a 522-fs pulse is measured. OSA: optical spectral analyzer, SMF: single-mode fiber, EDFA: erbium-doped fiber amplifier.

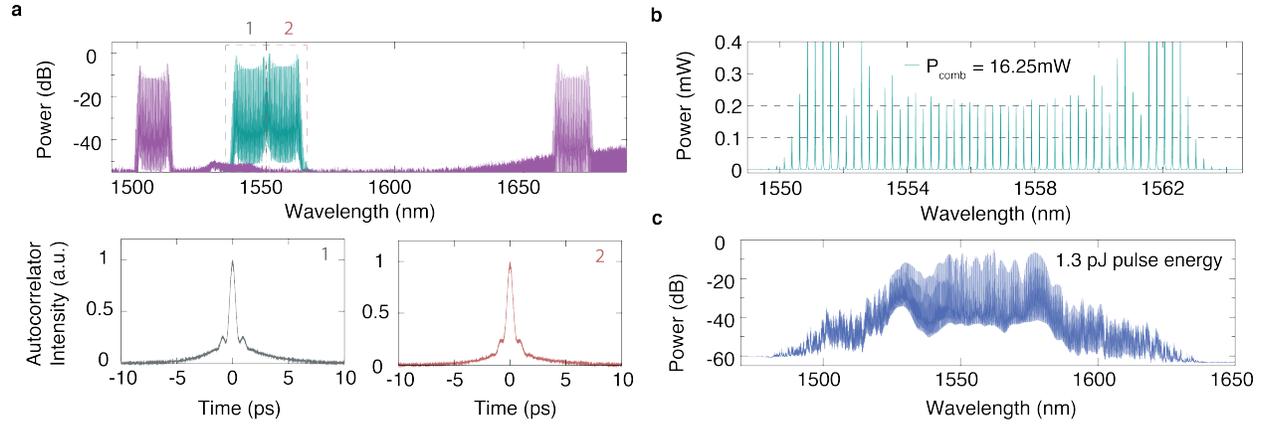

**Fig. 3 | Wavelength multiplexed, flat-top, and high power EO comb sources. a,** Spectrally-tailored comb spectra. Generation of two EO combs with different central wavelengths are obtained at the same time by pumping the LN chip with two combined CW lasers at user-defined wavelengths, at 1543.75 and 1556.83 nm (green) as well as at 1506.36 and 1669.51 nm (purple). Two individual modelocked pulse trains are generated simultaneously at a 30.135-GHz repetition rate, which correspond to EO-comb 1 and EO-comb 2, respectively (labeled in green spectrum). Both pulse trains emit ~520 fs pulses, which are individually verified by the autocorrelator traces (bottom). The combined pulses are also measured (see Methods). **b,** EO-comb power spectrum. The comb power is 16.25 mW at a pump power of 65 mW on chip, corresponding to a high pump-to-comb conversion efficiency of 25%. Both input and output facet loss are 3 dB. The spectral power is plotted in the linear scale and features a flat-top envelope. The central 25 comb lines have a power variation of < 1 dB. The highest comb-line power is 0.9 mW at spectral wings (see Methods). The power of 52 comb lines are above 0.1 mW, and 40 comb lines are above 0.2 mW. **c,** Nonlinear spectral broadening via a highly nonlinear fiber (HNLF). The 520-fs pulse train at a center wavelength of 1556.8 nm is softly amplified and sent to a 1-km HNLF at an average power of 40 mW and a pulse energy of 1.3 pJ. The final spectral bandwidth is 150 nm, broadened from an initial bandwidth of 16 nm (-60-dB level).

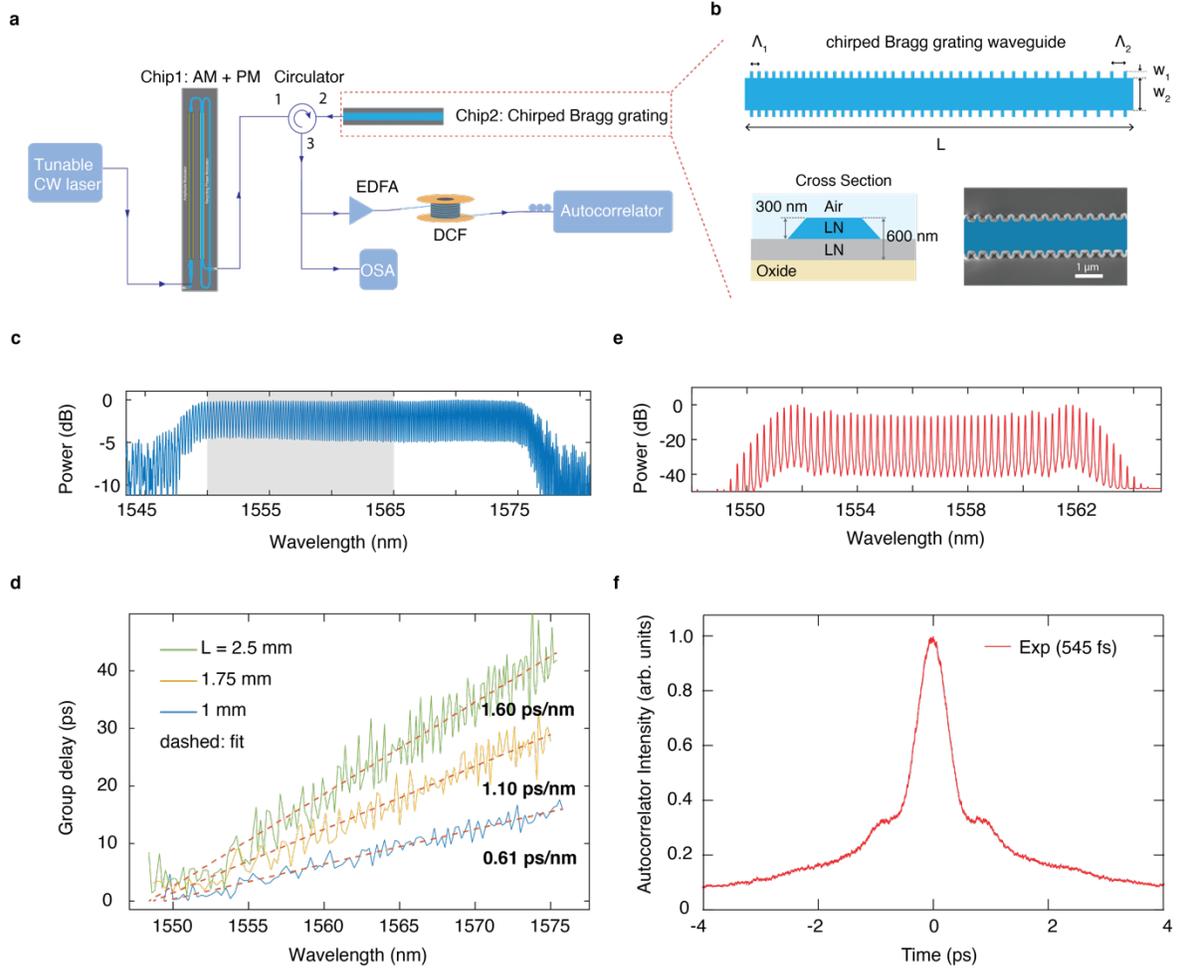

**Fig. 4 | Lithium-niobate-based Bragg chirped grating for on-chip pulse compression. a,** Experimental setup using chirped Bragg grating for pulse compression. **b,** Integrated chirped Bragg grating on thin-film LN (zoom-in of the chip 2 in (a), dashed box). The grating period is linearly chirped from $\Lambda_1 = 406.5$ nm to $\Lambda_2 = 414.5$ nm along the grating length $L$. The Bragg grating device is air cladded and fabricated on a 600-nm thin-film LN wafer with a fin width $w_1$ of 200 nm and a waveguide width $w_2$ of 1.1 μm. **c,** Reflection spectrum of the grating (top) characterized using a tunable CW laser. The grating 3-dB bandwidth is 1547-1576 nm, which covers the optical bandwidth of the time lens output (shaded). **d,** Group delay dispersion of three different grating lengths, extracted from the wavelength-dependent fringe periods in its corresponding reflection spectrum. The dispersion value is extracted to be 0.61, 1.10 and 1.60 ps/nm for grating length of 1, 1.75 and 2.5 mm, respectively. **e,** The optical spectrum of the time lens reflected after the grating. **f,** The autocorrelator trace of the pulse at the output of the grating. The pulse is fully compressed to its Fourier transform limit of 545 fs using a 2.25-mm integrated LN Bragg grating.

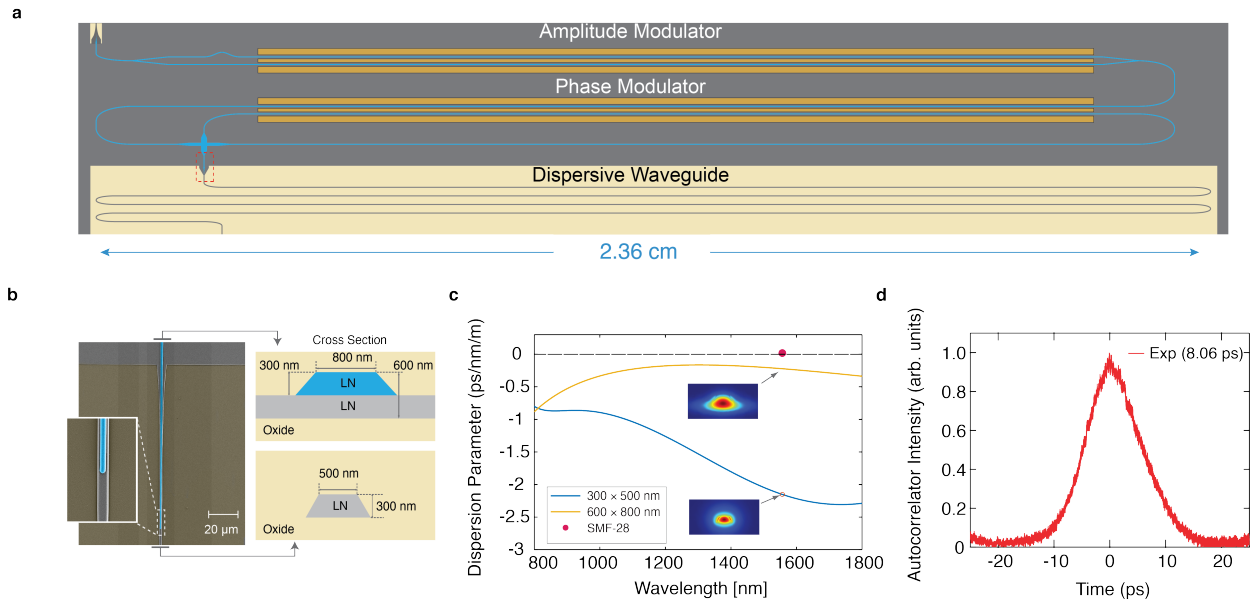

**Fig. 5 | Fully integrated time lens system on thin-film lithium niobate. a,** Integrated highly dispersive waveguide is used for pulse compression. The waveguide is patterned with a serpentine structure with a bending radius of 100 µm and a total length of 9.56 cm. This is smaller than ideal length of 49 cm and is limited by the waveguide losses of 300 nm thick compression waveguide of 1.5 dB/cm. **b,** SEM micrograph of the transition taper connecting the PM output with the compression waveguide. It starts with a partially etched waveguide with a top width of 800 nm. The top and slab width of the waveguide is adiabatically tapered to 300 nm and 1 µm over a length of 120 µm (left). The fully-etched slab waveguide is then slowly tapered to a final width of 500 nm over a length of 55 µm. **c,** The dispersion parameter $D$ of the integrated LN waveguides. The compression waveguide (300 × 500 nm) is dispersion engineered to have a large dispersion value of − 2.15 ps/nm/m at 1557 nm. Its absolute value is a factor of 10 and 120 larger than the |D| of the partially etched waveguide (600 × 800 nm, D = -0.22 ps/nm/m) and single-mode fiber (SMF-28, D = 0.018 ps/nm/m). Therefore, the total required length to reach full pulse compression at 30 GHz is 0.49 m, as compared to 59 m of SMF-28 in Fig. 2d. **d,** On-chip pulse compression down to a FWHM of 8.06 ps, which corresponds to the total GDD of the 9.56-cm waveguide.

## Methods

**Device fabrication, parameters, and optical loss characterization**

Devices are fabricated on a thin-film X-cut lithium niobate on insulator wafer (NanoLN) with a film thickness of 600 nm and an oxide insulator thickness of 2 μm. Electron-beam lithography (EBL) is used to define the optical waveguide with hydrogen silsesquioxane (HSQ) resist, which is then partially etched by Ar+ based reactive ion etching with an etch depth of 300 nm. The thickness of the remaining slab is 300 nm. The device is then cladded with silicon dioxide via plasma-enhanced chemical vapor deposition (PECVD). Photolithography and hydrofluoric acid wet etching is used to remove the oxide cladding in the regions where the spot size converter as well as the highly dispersive waveguide are located (Fig. 2a). A second EBL and reactive ion etching (RIE) process is used to define the final input taper and the compression waveguide on the slab layer, which is followed by PECVD oxide cladding. Microwave electrodes are patterned using both EBL and photolithography, and gold metal contacts are deposited using e-beam evaporation and lift-off processes. The facets of the chip are etched by deep RIE for end-fire optical coupling.

The width of partially etched optical waveguides is 1.5 μm, except that the width of the Y-splitter waveguides is 0.8 μm to avoid coupling to higher-order spatial modes. The cladding thickness is 820 nm. The dimension of the spot size converter at both facets is 300 × 250 nm for off-chip coupling. The coupling loss to the lensed fiber (mode size of 2 μm diameter) and distributed feedback laser chip is measured to be 3 dB and 2 dB, respectively. The propagation loss on chip is measured to be 3.2 – 3.6 dB, and ~0.3 dB/cm on the partially etched waveguides given a total length of 10 cm. The dimension of the crossed waveguide on the phase modulator is shown in Extended Data Fig. 1a, and its insertion loss is measured around 0.26 dB extracted from a linear fit of the loss on a separate chip with multiple crossings (Extended Data Fig. 1b). The microwave strip has a metal thickness of 1.6 μm, an electrode gap of 5 μm and a signal and ground pad width of 35 μm and 290 μm, respectively.

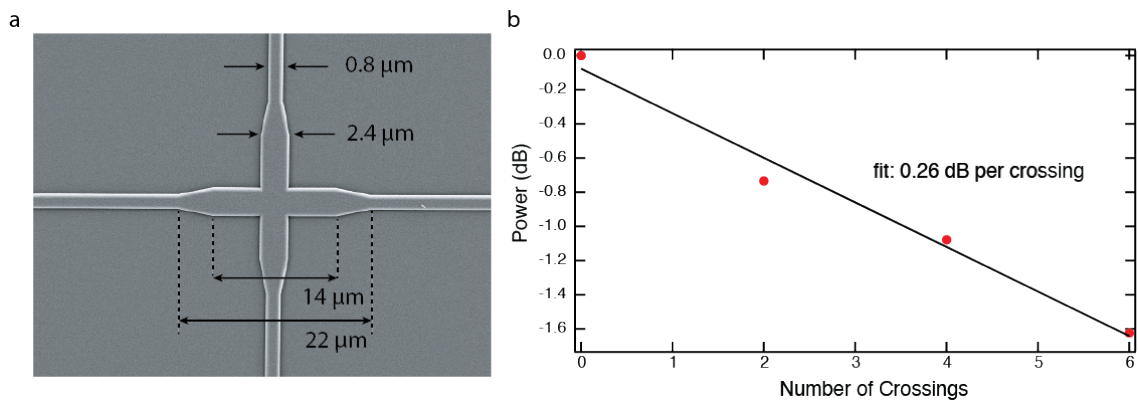

**Extended Data Fig. 1 | Illustration of the waveguide crossing in the phase modulator. a,** the geometry of the crossing on a partially etched waveguide with a 300 nm etch depth. **b,** the insertion loss extracted to be 0.26 dB per crossing.

**Inverse taper design and coupling loss characterization.** We designed the geometry of our inverse taper at the LN chip facet for efficient input coupling from the on-chip III-V laser, as shown in Extended Data Fig. 2 a & b. Our coupling experiments are performed on a 4 mm long LN waveguide with inverse tapers at each facet. We first measure the fiber-to-fiber loss using lensed fiber input to extract a coupling loss of the couplers from lensed fiber (= 3 dB). The input lensed fiber is then replaced by the DFB laser to measure the laser to spot-size converter loss. In the Extended Data Fig. 2c, we plot the transmission at the laser-LN interface as we sweep the air gap in between. We measure the lowest coupling loss of 2 dB at 1.8 µm gap, which matches our simulation results. In our simulation, an antireflective coating of $Al_2O_3$ at the laser interface is implemented to match the refractive index difference of the laser mode and air. The experimental coupling loss is limited by the gap between the laser-LN interface, caused by a combination of protruding silicon substrate and a slight angle between the laser and LN chip. We note that both the simulation and experimental results show similar periodic behavior in the transmission curve, which is due to the cavity formed in the air gap between the laser and LN facets.

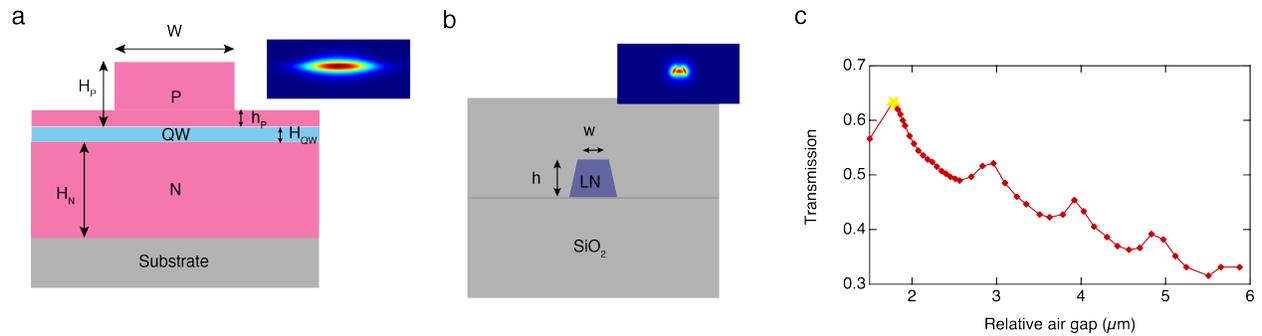

**Extended Data Fig. 2 | Geometric properties of the coupled structures, and the coupling loss characterization. a,** the DFB III-V laser waveguide where W = 5 um, $H_P$ = 2.5 um, $h_P$ = 500 nm, $H_{QW}$ = 450 nm, $H_N$ = 4 um. Inset: simulated mode profile at the laser output. **b,** the LN inverse taper where w = 250 nm, h = 300 nm. Inset: simulated mode profile. **c,** experimental results for coupling loss as a function of the air gap between the laser and LN chip facet. We operate at 63% transmission (2-dB coupling loss) at an air gap of 1.8 µm (labelled as star).

**Characterization of amplitude and phase modulators.** The microwave impedance is 37 Ω at 10 and 30 GHz. The measured microwave phase index is 2.27, which matches well with the optical group velocity index at 1.55 µm. We have measured the microwave loss of the 2-cm transmission line using a 50- Ω vector network analyzer (VNA) and have obtained 7.4 dB and 10.7 dB total loss at 10 and 30 GHz, respectively. The ohmic loss is characterized at 0.7 $dB/GHz^{1/2}/cm$.

The electro-optic response of the Mach-Zehnder interferometer (MZI)-based amplitude modulator (AM) is measured using a VNA, as shown in Extend Data Fig.3. A continuous-wave laser (Santec TSL 510) is coupled into the time lens chip. The AM is biased at the quadrature point. The output is softly amplified to 1 mW and sent to a high-speed photodetector (Newport 1014, 45 GHz bandwidth), followed by the VNA. The microwave power output from the VNA is at -20 dBm to avoid nonlinearities. After the calibration of the cable loss and detector response, the electro-optic (EO) $S_{21}$ response of the AM is plotted along with the theoretical prediction in the Extend Data Fig.3, indicating a 3-dB EO bandwidth > 45 GHz. The half-wave voltage ($V_\pi$), which is the voltage which induces a phase difference of π radian between the arms of the MZI, is measured to be 1.25 V at 1 kHz using a real-time oscilloscope.

The $V_\pi$ of the "recycling" phase modulator (PM) corresponds to the voltage which results in a total phase accumulation of $\pi$ radians in the PM, and is extracted from the optical spectra where the sideband power follows the Jacobi-Anger expansion:

$$E_0 e^{i2\pi f_c + i\beta \sin(2\pi f_{RF} t)} = E_0 e^{i2\pi f_c}(J_0(\beta) + \sum_{k=1}^{\infty} J_k(\beta)e^{ik2\pi f_{RF} t} + \sum_{k=-1}^{\infty}(-1)^k J_k(\beta)e^{-ik2\pi f_{RF} t})$$

where $E_0$ is the electric field amplitude, $f_c$ is the optical carrier frequency, $f_{RF}$ is the microwave frequency, $\beta$ is the modulation index ($= \pi \times V/V_\pi$ radian), $J_k(\beta)$ is the $k^{th}$ order Bessel function. The RF power is set at $\beta < 0.75\ \pi$ so that the optical power of the pump line decreases as the RF power increases. The $V_\pi$ is fit based on $\frac{P_0(\beta=V)}{P_0(\beta=0)} = |J_0\left(\pi \frac{V}{V_\pi}\right)|^2$ as we sweep the RF frequency from 4 to 40 GHz (Extended Data Fig. 4a). The RF frequency range is limited by the resolution of the optical spectral analyzer (> 2 GHz) and the range of the signal generator (< 40 GHz). The extracted $V_\pi$ is plotted in Extended Data Fig. 3b, where the lowest $V_\pi$ is 2 - 2.6 V from 4.4 to 38.7 GHz. Related to our operating conditions in the main texts, the $V_\pi$ is 2.2 and 2.5 V at 10.075 and 30.135 GHz, respectively. Extended Data Fig. 4c shows a 3-dB RF power bandwidth of 1.4 V at 30.1 GHz, which is 50% of the RF frequency period of 2.84 GHz (Extended Data Fig. 4d). The optical delay length of the LN waveguide between the first ground-signal (GS) electrode pair to the second GS pair is 46.5 μm, which results in an optical delay time of 0.35 ps ($\approx \frac{46.5\ \mu m}{c/n_0}$, where $c$ is the light velocity, $n_0$ is the optical group index of 2.27). The RF frequency period (free spectral range) is 1/optical delay time (2.84 GHz), which agrees well with the experimental measurement. Finally, the resonant $V_\pi$ is plotted in Extended Data Fig. 4e, along with a theoretical calculation that takes into account all the optical and RF properties of the fabricated device.

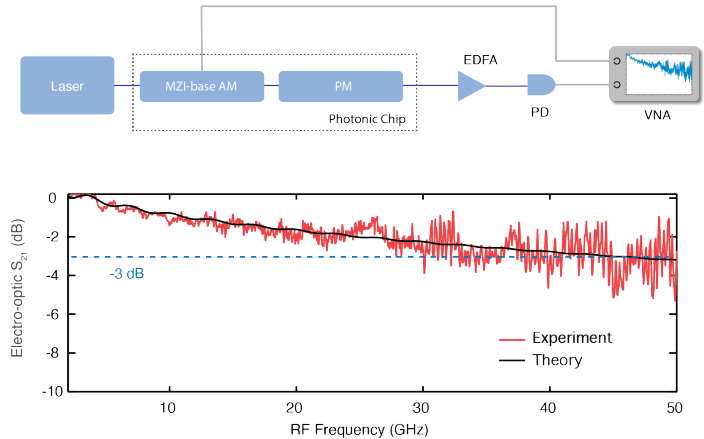

**Extended Data Fig. 3 | The electro-optic response of the amplitude modulator on the time-lens chip.** The response is measured and calculated based on the $S_{21}$ of the VNA and the calibration of the fast photodetector and RF cables, when the AM is biased at the quadrature point and under small signal modulation (top). The EO $S_{21}$ response of the AM indicates a 3-dB bandwidth of > 45 GHz (bottom). Since the impedance mismatch results in a shape fall off in the response at near

DC frequencies, the EO $S_{21}$ is often referenced to an RF frequency (2 GHz in our case). MZI: the Mach-Zehnder interferometer; AM: amplitude modulator; PM: phase modulator; EDFA: erbium-doped fiber amplifier; PD: photodetector; VNA: vector network analyzer.

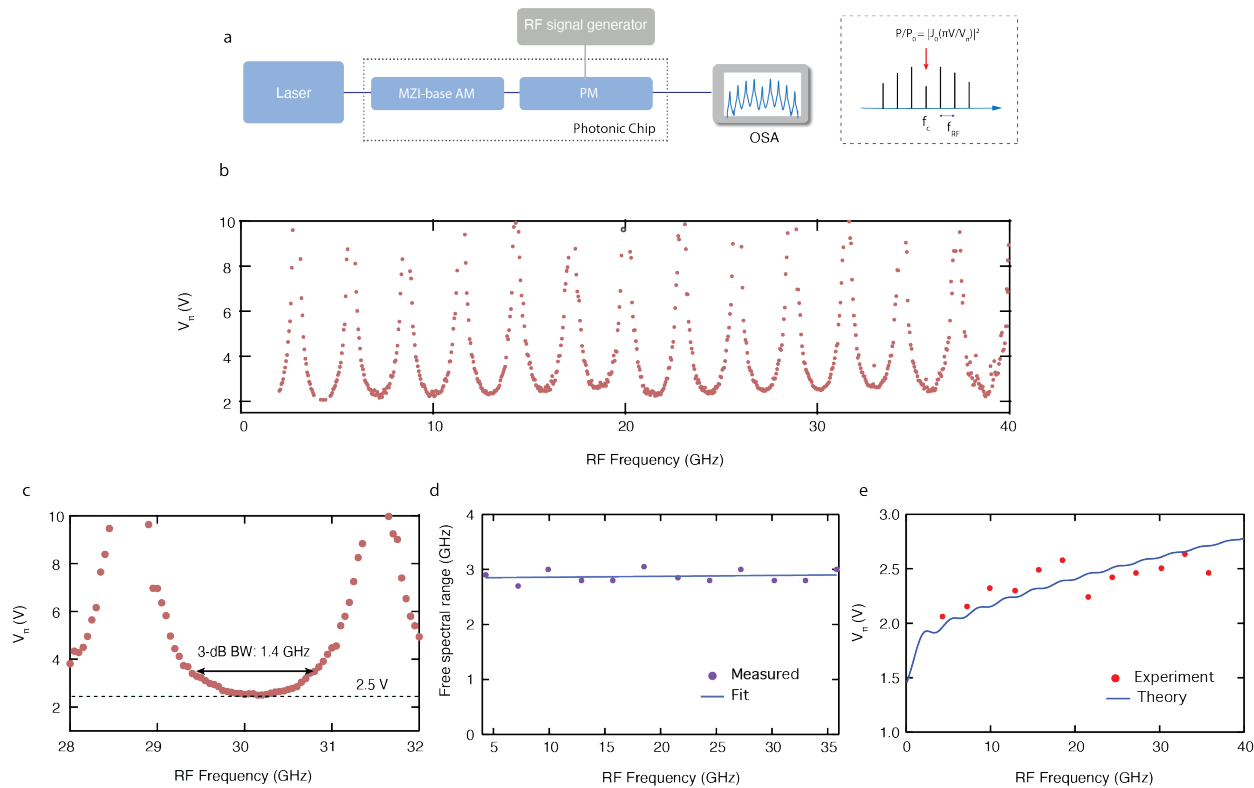

**Extended Data Fig. 4 | Characterization of the half-wave voltage ($V_\pi$) of the "recycling" phase modulator. a,** the experimental setup for characterizing the $V_\pi$. The optical spectra are recorded as we sweep the RF frequency from 2 to 40 GHz at a step of 50 MHz. $V_\pi$ is extracted based on the pump power depletion in the optical spectrum, which is determined by the magnitude square of the zero-order Bessel function $|J_0(\beta)|^2$, where $\beta = \pi \times V/V_\pi$ and $V$ is the driven voltage. **b,** the extracted $V_\pi$ as a function of the RF frequency. **c,** the $V_\pi$ around 30.1 GHz, which shows the lowest $V_\pi$ of 2.5 V at 30.1 GHz. The 3-dB power bandwidth, corresponding to $\sqrt{2}V_\pi$, is 1.4 GHz (29.4 to 30.8 GHz). **d,** the frequency period of the $V_\pi$ (free spectral range = 2.84 GHz). It is determined by $1/\tau$, where $\tau$ is the optical delay time from the first ground-signal (GS) pair to the second GS pair in the phase modulator. **e,** the $V_\pi$ at each resonant RF frequency. In the experiment, we operate at 10.075 and 30.135 GHz which corresponds to 2.2 and 2.5 V, respectively. The results are in good agreement with the theory, which is based on the measured RF properties (index, loss, impedance) and optical group refractive index. OSA: optical spectral analyzer.

**Numerical simulation.** The time-lens output in the frequency and temporal domain can be modelled numerically based on a sequential operation of amplitude modulation, phase modulation and dispersion operator. Starting from a continuous-wave wave ($E_0$), the field $E$ become

$\frac{E_0}{\sqrt{2}}(e^{j\frac{\pi}{4}\cos(2\pi f_{RF})} + e^{-j\frac{\pi}{4}\cos(2\pi f_{RF})+j\frac{\pi}{2}})$ after driving an amplitude modulator biased at the quadrature point with a peak-to-peak voltage equal to $V_{\pi,AM}$. Next, the phase modulator induces a phase modulation term $e^{j\frac{\pi V_{pp}}{V_{\pi,PM}}\cos(2\pi f_{RF}+\phi_0)}$, where $V_{pp}$ is the peak-to-peak voltage at the PM and $\phi_0$ is the RF phase difference between the AM and PM field. At last, the field is sent to a dispersive medium which imparts a phase $e^{-j\frac{1}{2}\beta_2(2\pi f - 2\pi f_c)^2 L}$ in the frequency domain where $\beta_2$ is the group velocity dispersion and $L$ is the length of the dispersive medium. The simulated optical spectrum and the temporal pulses carved by the AM before and after compression are shown in Extended Data Fig. 5, where the experimental parameters are used ($f_{RF} = 30.135$ GHz, $V_{pp} = 7.8\, V_{\pi,PM}$, $\phi_0 = \pi$, $\beta_{2,SMF} = -2.2958 \times 10^{-26}$ s$^2$/m, $L = 59$ meters). The optical spectrum agrees well with the experimental results in Fig. 2d, and the compressed pulse gives a full-width-half-maximum (FWHM) of 526 fs, as compared to the initially AM-carved pulse FWHM of 16.56 ps. The intensity autocorrelator trace can be simulated using the equation $I_{auto}(\tau) = \int_{-\infty}^{+\infty} I(t)I(t-\tau)dt$, shown in Fig. 2d.

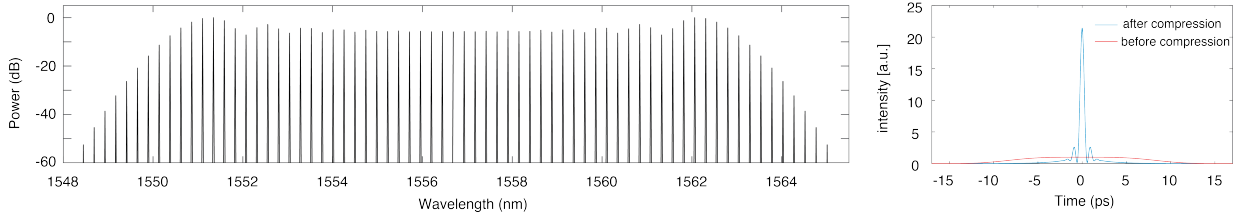

**Extended Data Fig. 5 | Numerical simulation of the optical spectrum and the temporal pulses of the time-lens output.** The optical spectrum (left) shows a 10-dB optical bandwidth of 12.5 nm, and the temporal pulse shows a FWHM of 526 fs after compression (right), which agrees excellently with the experimental results in Fig. 2d at 30.135 GHz. In the simulation, the dispersion medium used is the single-mode fiber (SMF-28) with a length of 59 meters. The temporal-domain behavior shows a pulse compression factor of 31.5 before and after the dispersion operator.

**Characterization of wavelength multiplexed pulse trains.** The intensity autocorrelation of two wavelength-multiplexed pulse trains at center wavelengths of 1543.75 and 1556.83 nm are shown in Extended Data Fig. 6. Each pulse train is verified and shown in Fig. 3a by blocking the other CW laser. The combined trace shows the feature of one major peak (B) with two weaker peaks on both sides (A & C) within one temporal period (1/30.135 GHz ≈ 33.18 ps). Based on the dispersion of the SMF-28 fiber of ~18 ps/(nm·km) at 1550 nm, the temporal delay between the two pulses after 59 meters of fibers is calculated to be 13.8 ps (= 18 ps/(nm·km) × 0.059 m × (1556.83-1543.75) nm). Therefore, the two weaker peaks in the autocorrelator trace correspond to the cases where only one pulse train overlaps in the temporal domain (A & C) while the highest peak results from both pulse trains overlaps with the copies of themselves (B). The temporal delay can be extracted experimentally from the trace as well, which is 13.3 ps in our case. The deviation from the numerical calculation is a result from the calibration of the autocorrelator and the accuracy of the estimated length and the dispersion value of the SMF-28 fibers.

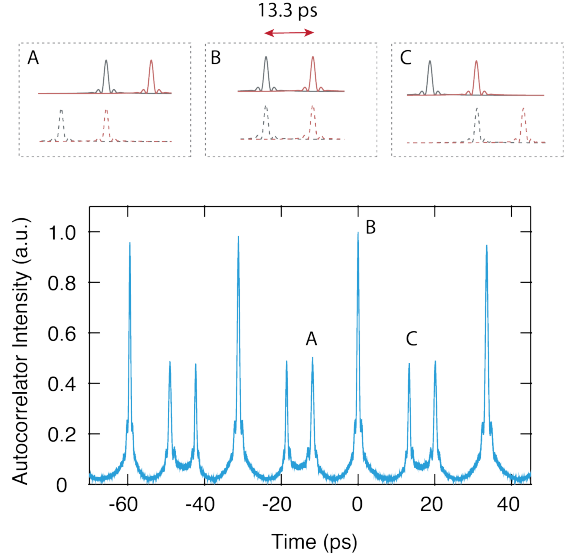

**Extended Data Fig. 6 | Autocorrelator trace of wavelength multiplexed pulse sources.** The experimental trace is obtained from the intensity autocorrelator when two pulse trains are simultaneously synthetized using two CW lasers at wavelengths of 1543.75 and 1556.83 nm, corresponding to Fig. 3a.

**Nonlinear measurements using highly nonlinear fibers (HNLF).** The femtosecond pulses generated from the time-lens chip at 1556 nm are amplified to an average power of 80 mW, and then spectrally filtered using a band-pass filter with a passband between 1543 nm and 1565 nm (Extended Data Fig. 7). The amplified pulses are then sent to a 1-km-long highly nonlinear fiber (HNLF), which has a fiber-to-fiber coupling loss of 3 dB. The input for nonlinear broadening is a 30-GHz pulse train with a pulse duration (FWHM) of 520 fs and a peak power of 2.5 W. The HNLF has a zero-group-velocity-dispersion (GVD) wavelength of 1549 nm with a slope of 0.016 $ps/nm^2 \cdot km$. The nonlinear coefficient is 10 $(W \cdot km)^{-1}$. The broadest spectrum is obtained at near-zero GVD point and GVD < 0 (Fig. 3c). The broadened comb linewidth is observable in the recorded optical spectrum below 1510 nm and beyond 1600 nm, which is a result of partially incoherent broadening. The spectra at wings are partially generated by the modulation instability from the noise. The spectral coherence can be improved by further reducing the pulse duration as well as a more stabilized microwave generator to reduce the cascaded phase noises.

In addition to nonlinear broadening in HNLF, we perform a four-wave mixing (FWM) experiment by combining the time-lens source with a tunable continuous-wave (CW) laser as the probe. The Extended Data Fig. 8 shows the optical spectra when the wavelength of the CW laser is set to be 1600, 1620, 1640, 1660, and 1680 nm, where combs are generated around the probe laser wavelength due to cross-phase modulation from the pump as well as at the idler wavelength from the FWM process. The broadest idler bandwidth achieved is 40 nm (from 1457 to 1497 nm) for a CW laser wavelength of 1640 nm. Both experiments show the potential of using the high-peak power pulses from the photonic-chip-based time-lens source as a pump for nonlinear optical frequency conversion. One can envision replacing the HNLF with integrated LN or periodically-poled LN waveguides or microresonators, as a fully compact photonic circuit, for efficient $\chi^{(2)}$ or $\chi^{(3)}$-based parametric processes.

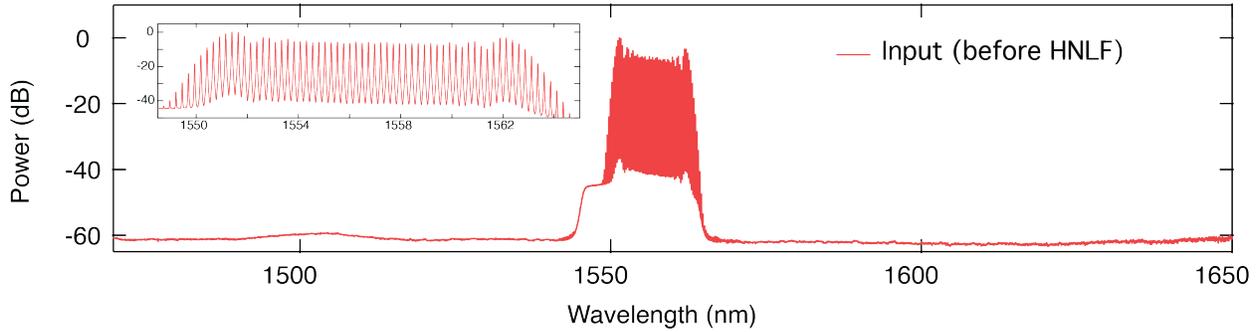

**Extended Data Fig. 7 | Input optical spectrum for nonlinear broadening in highly nonlinear fiber (HNLF).** The time-lens output pulses are softly amplified to 2.6 pJ pulse energy using an erbium doped fiber amplifier (EDFA), and subsequently filtered using a band-pass optical filter (passband range: 1543-1565 nm) to filter out the amplified spontaneous noises (ASE). The pedestal in the optical spectrum is a result of the remaining ASE from the EDFA. The coupling loss to HNLF is 3 dB, which results in a pulse peak power of 2.5 W for nonlinear broadening (shown in Fig. 3c). Embedded: zoom-in spectrum with a center wavelength of 1556 nm.

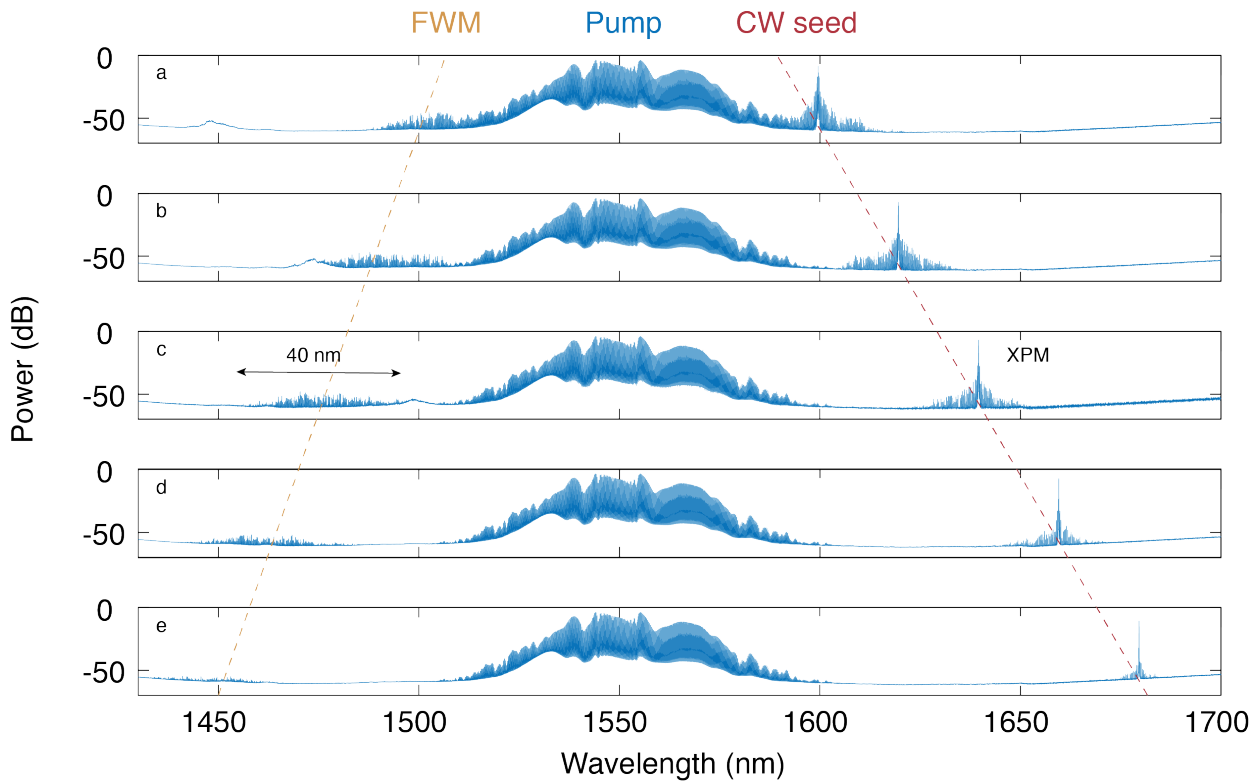

**Extended Data Fig. 8 | Four wave mixing in HNLF using the pulse train as the pump and a CW laser as the probe.** Five optical spectra are recorded when the time-lens pulse train at 1549 nm is sent to the HNLF along with a continuous-wave (CW) laser at wavelengths of 1600, 1620, 1640, 1660, and 1680 nm (a-e), respectively. In addition to the nonlinear broadening around the

pump wavelength, mini-comb spectra are generated around the wavelength of the CW laser due to the cross-phase modulation (XPM) from the high-peak-power pulses. Four-wave mixing between the pulse pump and the CW laser probes results in broad comb spectra at the idler wavelength on the lower wavelength side of the pump. The broadest conversion bandwidth is around 40 nm when the CW laser is at 1640 nm.

**Experimental set-up and characterization for on-chip pulse compression.**
**Chirped Bragg grating.** The grating device is first characterized using a tunable CW laser and a fiber-based circulator (Extended Data Fig. 9). Here we use a 2.5-mm long grating as an example. The input polarization is set at the TE mode using a polarization controller. Both reflection and transmission are recorded using a photodetector while laser frequency is swept. The fringe in the reflection spectrum is due to the cavity effect from reflection between the input facet and the grating. Since we oriented the grating such that the grating period linearly increases in the input light direction, the fringe period (in nm) gradually decreases as the effective cavity length gets longer at a longer wavelength. The transmission spectrum shows an extinction ratio of 10 dB, which is limited by our detector noise floor (-18dBm). We further increase the power of the input light by 13 dB (maximum power) and observe the transmission inside the bandgap remains at the noise floor level. It indicates an extinction ratio > 23 dB, which corresponds to > 99.5% reflectivity (< 0.02 dB loss). Extended Data Fig. 9d plots the fringe period ($\Delta\lambda$) as a function of wavelength, which can be used to extract the group delay (= $\lambda^2/c\Delta\lambda$) shown in Extended Data Fig. 9e. The propagation loss inside the grating is extracted using the input laser spectrum and the reflection spectrum. Since light of different wavelength has different propagation distance in the grating, we can convert the wavelength-dependent loss to propagation loss. Extended Data Fig. 9f shows a fitted loss to be $0.033 \pm 0.0017$ dB/mm. This is the lowest insertion loss reported in LN-chip grating.

To achieve full pulse compression using grating waveguide (main text Fig.4), we use a 2.25-mm long grating device which is oriented to operate at normal GVD regime. This facet that it is longer than the targeted grating length of 1.8 mm (corresponds to $7.6\pi$ modulation depth) is that we have an extra 16-meter of SMF-28 optical fibers in the optical path, which we need to compensate off with a longer grating length.

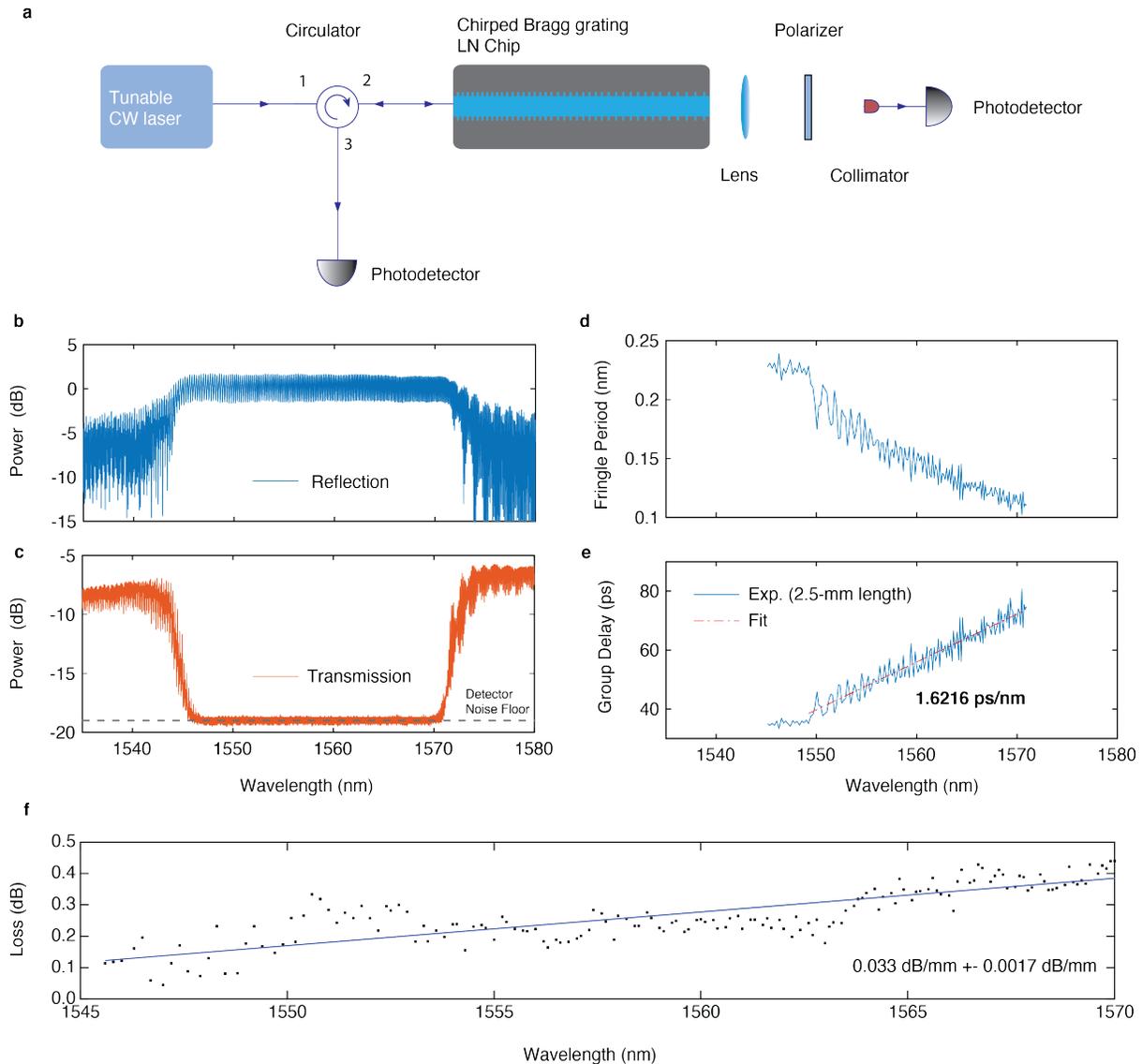

**Extended Data Fig. 9| Characterization of the on-chip chirped Bragg grating. a,** experimental set up. The device is measured using a tunable CW laser and a circulator. **b & c,** the reflection and transmission spectrum of a 2.5-mm Bragg grating. The transmission within the bandwidth is limited by our detector noise floor, instead of the device. **d,** the extracted fringe period from (b). The fringe is a result of the Fabry Perot cavity effect due to the reflection between input facet and grating. **e,** the group delay as a function of wavelength. The slope indicates a total dispersion value of 1.62 ps/nm. **f,** the insertion loss as a function of wavelength, extracted from the reflection spectrum and laser input power spectrum. Since the wavelength can be mapped to the physical propagation distance in the grating device, the propagation loss is fitted to be 0.033 dB/mm.

**Dispersive waveguide.** To characterize the temporal profile exactly at the chip output, we insert a 70-m-long dispersion compensating fiber (DCF4, Thorlabs) with a negative dispersion parameter of -4 ps/nm·km to compensate the total dispersion induced by the SMF-28 fibers and the fibers in the EDFA in front of the autocorrelator. The absolute value of the accumulated dispersion of the fiber link between the chip output and the autocorrelator is less than $1.0\times 10^{-26}$ s² (equivalent to the dispersion of < 0.5 m of SMF-28 fiber), which is calibrated based on the temporal broadening of a pulse laser source (Calmar Laser FPL-02CFF, 1.6-ps pulse duration) sent to the fiber link. In our experiment, this gives an error bar of ± 0.11 ps to our measured pulse duration of 8.06 ps at the chip output.

In order to achieve the shortest pulses after the on-chip compression, we drive the AM at an RF frequency of 15.07 GHz with a $V_{pp}$ of $2 \times V_{\pi,AM}$ while biasing at the maximum transmission point. Part of the RF generator signal is sent to a microwave frequency doubler which is followed by the RF amplifier. Based on this scheme, Extended Data Fig.10 plots the simulated pulse width as a function of the highly dispersive waveguide length. Given the experimental condition with a 9.56-cm-long compression waveguide (Fig.5), it is predicted to generate a pulse width of 8.03 ps at the chip output, which is in good agreement with the measured pulse width of 8.06 ps ± 0.11 ps.

The pulse compression can be achieved using a dispersive medium with either normal GVD and $\phi_0 = 0$ or anomalous GVD and $\phi_0 = \pi$, where $\phi_0$ is the microwave phase difference between the AM and the PM. Flipping the microwave phase difference between 0 and $\pi$ can be understood as changing the sign of the focal length of a lens (concave or convex). The on-chip waveguide has normal GVD while the SMF-28 fiber has anomalous GVD. In the Extended Data Fig.11, we configure the set-up to achieve the final pulse compression using SMF-28 fibers with an anomalous GVD, which requires an additional 11-meter fiber length to compensate for the normal dispersion of the on-chip waveguide.

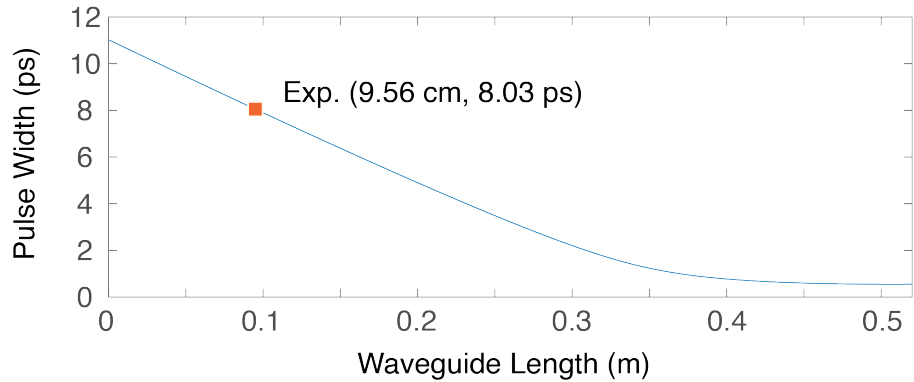

**Extended Data Fig. 10| The simulated pulse width as a function of the dispersive waveguide length.** The simulation is based on the experimental configuration where the AM is driven at half of the comb line spacing at 15.07 GHz and the PM is driven at 30.14 GHz. The GVD of the on-chip dispersive waveguide is + 2770 ps²/km at 1557 nm. A pulse width of 8.03 ps is obtained with a 9.56-cm-long waveguide based on the actual device parameter, which agrees well with the measured pulse width of 8.06 ± 0.11 ps shown in Fig. 5d.

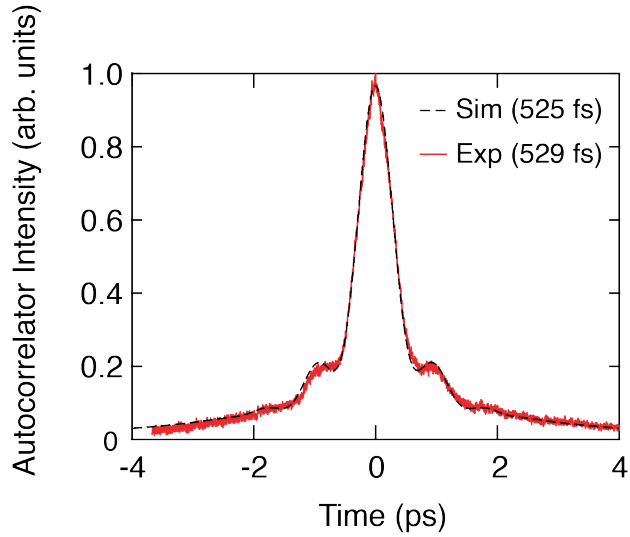

**Extended Data Fig. 11| The autocorrelator trace.** The final pulse duration of 529 fs, after compression with 70 meters of SMF-28. 11 meters of additional fiber is needed to compensate for the GVD of the compression waveguide, which is the opposite sign. The results agree well with the numerical simulation of the full system (dash curve).

**Table 1: Performance comparison with literature.**

| Platform | Pulse width | 10-dB Optical bandwidth | Repetition rate | Insertion loss | RF power (PM) | Stages of PM | Modulation index | 3-dB EO bandwidth | Ref |
|---|---|---|---|---|---|---|---|---|---|
| **Integrated** | 530 fs | 12.6 nm | 30 GHz | 8 dB (3 dB on chip) | 3.8 W | 1 | $7.8\pi$ | 45 GHz | This work |
| **Integrated** | 564 fs | 12.1 nm | 27 GHz | 8 dB | 3.8 W | 1 | $8.1\pi$ | 45 GHz | This work |
| **Integrated** | 1.14 ps | 6.0 nm | 10 GHz | 8 dB | 6.0 W | 1 | $11.2\pi$ | 45 GHz | This work |
| Bulk | 1.7 ps | 4.0 nm | 10 GHz | 12 dB*** |  | 2 | $7\pi$ | N/A | Ref 26 |
| Bulk | 600 fs | 11 nm | 10 GHz | 16 dB*** | 12 W | 3 | $21\pi$** | N/A | Ref 1 |
| Bulk | 600 fs | 11.5 nm | 30 GHz | 20 dB*** | 16 W | 4 | $7.1\pi$** | N/A | Ref 1 |
| Bulk | 984 fs | 5.2 nm | 12 GHz | 15-17 dB | 4 W | 3 | $10\pi$** | 20 GHz | Ref 27 |
| Bulk | 250 fs | 20 nm | 25 GHz | 17.2 dB* | 5-9 W | 3 | $17\text{-}18\pi$ | N/A | Ref 28 |

*PM mentioned for 4.3 dB insertion loss
** Extracted based on the optical bandwidth
*** Based on that commercialized LN modulator typically has at least 4dB insertion loss